\documentclass[12pt,letterpaper]{article}   

\usepackage[symbol]{footmisc}

\usepackage{graphics,graphicx}
\usepackage{times,color,url,amsmath}
\usepackage{float,psfrag}
\usepackage{tabularx}
\usepackage{latexsym,amsmath,amssymb}
\usepackage{verbatim}
\usepackage{setspace}
\usepackage{wrapfig}
\usepackage{multicol}
\usepackage{enumitem}
\usepackage{mathtools}
\usepackage{graphicx}
\usepackage{amssymb}
\usepackage{amsmath}
\usepackage{ulem}
\usepackage{epstopdf}
\usepackage{color}
\usepackage{graphicx}
\usepackage{xcolor}
\usepackage[letterpaper,top=1.00in, bottom=1.00in, left=1.00in, right=1.00in]{geometry}
\usepackage[super]{natbib}
\usepackage[backref]{hyperref}
\hypersetup{unicode=true,colorlinks=true,linkcolor=blue,citecolor=blue,filecolor=cyan,urlcolor=magenta}
\usepackage[labelfont={bf,stretch=0.85},textfont={stretch=0.85,it},aboveskip=0.35cm,belowskip=-0.15cm]{caption}

\usepackage{bbding}

\newcommand{\chandra}{{\sl Chandra}}
\newcommand{\rxte}{{\sl RXTE}}

\newcommand{\suzaku}{{\sl Suzaku}}
\newcommand{\xmm}{{\sl XMM-Newton}}

\newcommand{\swift}{{\sl Swift}}
\newcommand{\integral}{\textit{INTEGRAL}}
\newcommand{\nustar}{\textit{NuSTAR}}

\newcommand{\fermi}{{\sl Fermi}}

\everypar{\looseness=-1}
\usepackage[compact]{titlesec}
\titleformat*{\section}{\large\bfseries}
\titleformat*{\subsection}{\normalsize\bfseries}
\titleformat*{\subsubsection}{\normalsize\bfseries}
\titleformat*{\subparagraph}{\normalsize\bfseries}
\titlespacing{\section}{0pt}{0.28cm}{0.12cm}
\titlespacing{\subsection}{0pt}{0.23cm}{0.12cm}
\titlespacing{\subsubsection}{0pt}{0.1cm}{0.05cm}
\titlespacing{\subparagraph}{0pt}{0.08cm}{0.0cm}

\begin{document}
\pagenumbering{gobble}

\hrule
\vspace{0.5cm}
\noindent
{\sc \large Astro2020 Science White Paper}
\vspace{0.5cm}
\hrule

\vspace{0.65cm}
\noindent
{\Large \bf Magnetars as Astrophysical Laboratories of Extreme\\ 
Quantum Electrodynamics: The Case for a Compton Telescope}
\vspace{0.5cm}



\noindent {\bf Thematic Areas:} 
\hspace*{10pt} $\square$ \hspace*{-14pt} \CheckmarkBold Formation and Evolution of Compact Objects

\noindent
\hspace*{99pt} $\square$ \hspace*{-14pt} \CheckmarkBold Cosmology and Fundamental Physics

\vspace{10pt}\noindent
\textbf{Principal Author:} Zorawar Wadiasingh\\
\noindent
Institution:  NASA Goddard Space Flight Center, Gravitational Astrophysics Laboratory \\
Email: {\tt zorawar.wadiasingh@nasa.gov}\\
Phone:  {\tt +1 301 286 7063}\\
 \linebreak

\vspace{-20pt}\noindent
\textbf{Co-authors:}\\
\noindent
George Younes, George Washington University \\ 
Matthew G. Baring, Rice University \\ 
Alice K. Harding, NASA/GSFC \\
Peter L. Gonthier, Hope College \\
Kun Hu, Rice University\\ 
Alexander van der Horst, George Washington University \\ 
Silvia Zane, University College London \\ 
Chryssa Kouveliotou, George Washington University \\
Andrei M. Beloborodov, Columbia University \\
Chanda Prescod-Weinstein, University of New Hampshire  \\
Tanmoy Chattopadhyay, Pennsylvania State University \\
Sunil Chandra, North-West University Potchefstroom \\
Constantinos Kalapotharakos, NASA/GSFC \\
Kyle Parfrey, NASA/GSFC \\
Harsha Blumer, West Virginia University \\ 
Demos Kazanas, NASA/GSFC \\
\linebreak

\vspace{0cm}\noindent
\textbf{Abstract:}  
A next generation of Compton and pair telescopes that improve MeV-band detection sensitivity by more than a decade beyond current instrumental capabilities will open up new insights into a variety of astrophysical source classes.  Among these are magnetars, the most highly magnetic of the neutron star zoo, which will serve as a prime science target for a new mission surveying the MeV window.  This paper outlines the core questions pertaining to magnetars that can be addressed by such a technology.  These range from global magnetar geometry and population trends, to incisive probes of hard X-ray emission locales, to providing cosmic laboratories for spectral and polarimetric testing of exotic predictions of QED, principally the prediction of the splitting of photons and magnetic pair creation.  Such fundamental physics cannot yet be discerned in terrestrial experiments. State of the art modeling of the persistent hard X-ray tail emission in magnetars is presented to outline the case for powerful diagnostics using Compton polarimeters. The case highlights an inter-disciplinary opportunity to seed discovery at the interface between astronomy and physics.

\pagebreak

\pagenumbering{arabic}
\setcounter{page}{1}
\section{Magnetars In a Nutshell}


\noindent
Neutron stars serve as useful laboratories to study physics under
conditions of extreme density, gravity, and magnetic fields. Magnetars
represent a topical subclass of the neutron star family. The known
magnetars \citep{2014ApJS..212....6O} of our galaxy possess the
longest spin periods among all isolated neutron stars, yet with large
spin down rates. These temporal properties imply that they are young,
with an average spin down age of a few thousand years, possess the
highest magnetic fields in the Universe, with polar surface values of
$B_p \sim 10^{13}-10^{15}$~G, and exhibit weak spin-down power compared to
their more numerous ``cousins,'' the canonical rotationally-powered pulsars.

Magnetars spend much of their time in a quiescent state, where they
are observed as persistent quasi-thermal hot X-ray emitters with $k T \sim 0.5$ keV. Tellingly, their luminosities exceed their spin-down power by as much as three
orders of magnitude. Accordingly, magnetars cannot be powered by spin energy loss, but instead extract their power from the immense reservoir of magnetic energy, $10^{46} - 10^{48}$ erg. They occasionally enter burst active episodes where they emit a few to hundreds of short ($\sim 0.1$\,s), bright bursts in the 5--500 keV band
with $L_{\gamma}\sim 10^{37}-10^{42}$~erg~s$^{-1}$. Following the onset of such bursting
activity, magnetars enter an excited X-ray state where their quiescent
flux increases by factors ranging from a few to $1000$ times the
quiescent flux \citep{zelati18MNRAS:magOut,2017ApJS..231....8E}, phases named
``magnetar outbursts". These phenomena are usually accompanied by strong
spectral and temporal variations, e.g., hotter effective temperature,
glitch and anti-glitch events, strong timing noise, and pulse profile evolution\citep[][]{woods04ApJ:1E2259,israel07ApJ:1647,archibald13Natur,
  rea13ApJ:0418,kaspi14ApJ:1745,zelati15MNRAS:1745,scholz14ApJ:1822,
  alford16ApJ:1810,younes17:1935}. The outbursts may persist for
months to years, during which the magnetar spectral and temporal
properties recover to their
pre-outburst behavior\citep{zelati18MNRAS:magOut,2017ApJ...851...17Y}.

Despite the relatively low number of magnetars (23 confirmed, 6 candidates), they
possess an enormous topicality, as evidenced by the shear number of
dedicated reviews in the last 10 years\citep[][]{mereghetti08AARv:magentars,rea11:outburst,
  turolla15:mag, mereghetti15:mag,kaspi17:magnetars,esposito18:rev}.
Moreover, magnetars have been invoked to explain some of the extreme
phenomena in the Universe, such as super-luminous supernovae\citep[][]{kasen10ApJ:SNMag,inserra13ApJ:SNMag}, gamma-ray bursts\citep[][]{wheeler00ApJ, rowlinson13MNRAS} (GRBs), ultra-luminous
X-ray sources\citep[][]{2014Natur.514..202B,israel17Sci:ULXmag,ekcsi15MNRAS:ulxmag,dallosso15MNRAS} (ULXs), and the mysterious Fast Radio Bursts\citep[][]{2014MNRAS.442L...9L,masui15Natur,2016ApJ...826..226K,metzger17ApJ,2017ApJ...843L..26B} (FRBs). In short, these fascinating
objects remain at the forefront of astrophysics curiosity for the
foreseeable future.

\subsection{Current Observational Status and Gaps}

\noindent
The last two decades have been the golden age for nascent magnetar
science. \swift-BAT and \fermi-GBM have enabled  the discovery of a
large number of magnetars through the detection and localization of short
magnetar-like bursts\citep[][]{horst10ApJ:0418, kennea13ApJ},
\swift-XRT and \rxte-PCA have permitted the detailed study of their
temporal and spectral changes during outbursts\citep[][]{woods07ApJ:1906, cotizelati18MNRAS}, and last but not
least \chandra\ and \xmm\ have deciphered the quasi-thermal nature of
their persistent soft ($0.5-10$~keV) X-ray emission in quiescence and
during outbursts. 

A remarkable discovery was reported in 2004\cite{2004ApJ...613.1173K}
of a new {\textit{persistent}} spectral component in
1E~1841$-$045 with \rxte\ HEXTE between $10-150$~keV, {\textit{100\% pulsed
at the highest energies}}. Similar detections followed for
1RXS~J170849.0$-$400910, 4U~0142$+$61, and 1E~2259$+$586 using HEXTE
and \integral\ IBIS ISGRI \citep{2006ApJ...645..556K,
  2008A&A...489..245D,2008A&A...489..263D}. The higher sensitivity of
\nustar\ enabled the detection of these hard X-ray tails in
fainter magnetars; currently there are 7 magnetars that exhibit
persistent hard tails during quiescence, and another 6 during outburst
\citep{2017ApJS..231....8E}. These hard X-ray tails exhibit spectra consistent with
power laws (PL) of photon index $\Gamma\approx1.0$, demanding
drastic spectral changes at $\sim 10$ keV. Moreover, they dominate the
energetics, with fluxes exceeding that of the soft components, often
by factors of 10 or more. These hard power laws do not exhibit a
break below $100-200$~keV, and in a few cases, \integral, CGRO-COMPTEL and 
{\it Fermi}-LAT upper limits at energies $300-1000$~keV imply that a break must exist
in this soft $\gamma$-ray energy band.

{\sl Our understanding of magnetar energetics is
    incomplete.} Given these upper limits, only a sensitive soft $\gamma$-ray observatory would
    ultimately uncover the peak energy of the magnetar spectral energy
    distributions, thereby determining their total persistent energy
    budget in quiescence and in outburst. Moreover, these
observations would result in key observable parameters, such as the
exact shape and energy of the high-energy cutoff and its variations
with rotational phase. Equipped with polarization capabilities, such
an observatory would also reveal the polarization degree and
position angle signatures of the hard X-ray emission from magnetars. These
key observables depend critically on the photon and particle
interactions in one of the most extreme environments in the Universe,
and may result in the first discovery of exotic quantum electrodynamic
(QED) physics long thought to be operating in proximity of not only magnetars but also pulsars. 

\subsection{Theory in Brief}

\begin{wrapfigure}{r}{0.24\textwidth}
\centering
\vspace{-12mm}
\includegraphics[width=0.2399\textwidth]{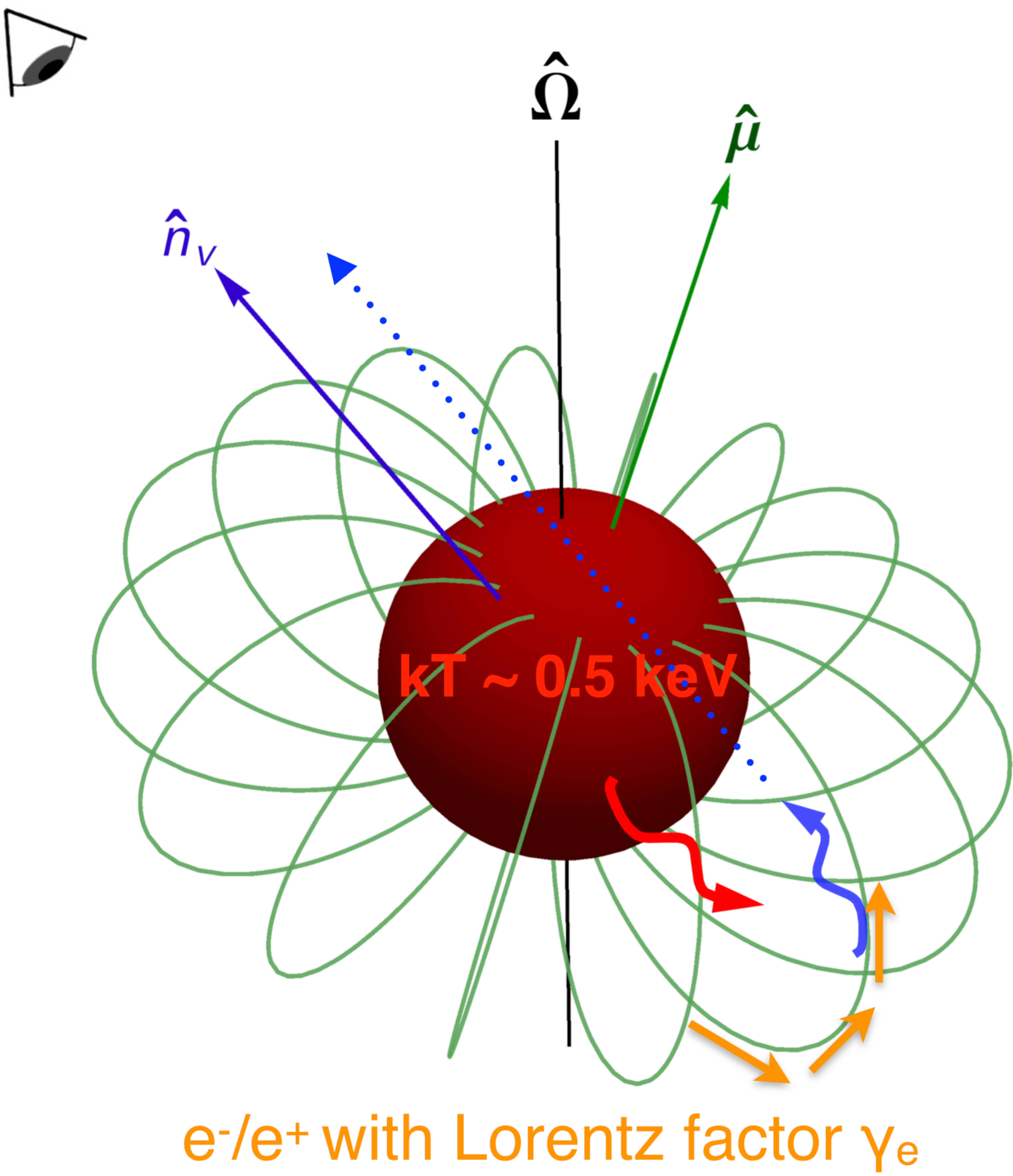}
\caption{In RICS, surface photons of energy $kT \sim 0.5$ keV are
  upscattered by relativistic electrons which follow field lines. The
  observer samples a small portion of the magnetosphere owing to the
  kinematics of RICS. For misaligned magnetic $\hat{\mu}$ and spin {\footnotesize $\hat{\Omega}$} axes, pulsations are
  observed.}
\label{cartoon}
\vspace{-5mm}
\end{wrapfigure}

\noindent 
The nonthermal nature of the persistent hard X-ray tails suggests that they
are powered by a relativistic electron/positron population.  Current
models are still in their infancy but steadily developing. In contrast
to normal pulsars, the persistent emission likely arises in the
``closed" zone of the magnetosphere where particle acceleration
proceeds in a magnetosphere that departs from ideal force-free 
magnetohydrodynamics. A quasi-equilibrium is established where particle acceleration,
pair production and radiative losses are in counterbalance
\cite{2007ApJ...657..967B,2013ApJ...777..114B}.

At low altitudes where emission likely originates, {{\sl resonant
    inverse Compton scattering (RICS)}} of the soft thermal surface
photons is the dominant radiative process for electrons that is
germane to the generation of hard X-ray tails
\cite{2007Ap&SS.308..109B,2007ApJ...660..615F,2008MNRAS.389..989N,2011AdSpR..47.1298Z,2011ApJ...733...61B,2013ApJ...762...13B,2014ApJ...786L...1H,2018ApJ...854...98W}. The
scattering cross section is greatly enhanced at the cyclotron
fundamental, where the incoming photon energy is equal to the
gyroenergy $\hbar \omega_B$ in the electron rest frame. For the
magnetar context, it is crucial to recognize that fields are in the
QED domain where $\hbar \omega_B \sim m_e c^2$; this defines the
critical field $ m_e^2c^3/(\hbar q_e)  \equiv B_{\rm cr} \approx
4.413\times 10^{13}$ G. Rapid cyclotron cooling restricts electrons to
move parallel to the field (see Fig~\ref{cartoon}).  Strong Doppler
beaming anisotropy and flux (and photon energy) boosting then result from RICS, which 
is imprinted on light curves, and traces the field geometry (electron
motion) and locales of the particles acceleration and cooling.  RICS produces a relatively flat spectrum, with high linear polarization degree, which cuts off at a
kinematically determined energy \citep{2018ApJ...854...98W}.


Magnetar magnetospheres are also opaque for hard X-rays and $\gamma$ rays.  The measured spectral cutoffs may also be produced by attenuation of photons principally due to {\sl{magnetic photon splitting ($\gamma + B \rightarrow \gamma\gamma$ )}} and {\sl{pair production ($\gamma + B \rightarrow e^+e^-$)}}. These exotic QED propagation effects \citep{1997ApJ...476..246H,2001ApJ...547..929B,2006RPPh...69.2631H} which are as yet untested terrestrially, imprint telltale polarimetric signatures on magnetar spectra and pulsations that can be probed with updated telescope technology.


\subsection{Questions That a Sensitive Compton Telescope Will Answer}

\noindent Extant models of the type discussed here provide an array of
possible spectral and polarization predictions that serve as a toolkit
for probing both geometry and physics of magnetars.
Accordingly, an array of important advances to our understanding of
these topical objects can be delivered with the deployment of a
mission with Compton detection technology that has both improved
continuum sensitivity and polarimetric capability above $100$ keV.
These deliverables include

\vspace{-2mm}
\begin{itemize}[leftmargin=*,labelsep=4mm]
\setlength\itemsep{-1mm}
\item employing variations of spectra and polarization with pulse
  phase to constrain the locale for hard X-ray tail emission -- Doppler
  boosting varies substantially for different sites of scattering.
\item phase-resolved spectroscopy to constrain the array of possible
  angles $\alpha$ between the rotational and magnetic axes of a
  magnetar.  This can be determined for a variety of magnetars, and
  trends of $\alpha$ with magnetar age can be explored.  Refinement of
  $B_p$ estimation then becomes possible. 
\item fundamental QED physics can be probed by ascertaining whether
  photon splitting and/or pair creation impose upper limits to the 
  emission energies.  Polarimetry enhances this diagnostic.
\item exploring activation of magnetar magnetospheres 
  following burst-active episodes relative to long-term relaxed
  conditions, thereby informing magnetar energetics and wind
  properties.
\end{itemize}
\vspace{-2mm}
Enabling these insights advances our understanding of
magnetars and their relationship to other neutron star varieties.
Yet the science reach extends to GRBs, ULXs and FRBs, each with
possible magnetar connections.  

\section{Details: State-of-the-Art Magnetar Models \& Pertinent QED Processes} 

\noindent
Soft X ray photon densities and magnetic field strengths are high at low altitudes, and so there the dominant
energy loss mechanism for electrons is RICS, which may be regarded as
cyclotron absorption followed by spontaneous re-emission, preserving
the electron in the ground Landau state. In the Thomson limit, the
maximum upscattered photon energy (in units of $m_ec^2$) is $\gamma_e (B/B_{\rm cr})\sim \gamma_e^2\epsilon_s$ while it is 
$\gamma_e$ in the Klein-Nishina regime, for electron Lorentz
factor $\gamma_e$, and surface thermal photon energy $\epsilon_s m_ec^2 \sim 0.1 -3$ keV. The
conditions for resonance are always satisfied in a thermal photon bath\cite{2011ApJ...733...61B}. In high $B\gtrsim B_{\rm cr}$ fields, a full QED treatment is necessary for cyclotron lifetimes, RICS cross sections and scattering kinematics \citep{2018ApJ...854...98W,2000ApJ...540..907G,2005ApJ...630..430B,2014PhRvD..90d3014G}. As in Thomson scattering, RICS generates distributions of photons with high {\textit{linear}}~polarization degree. The
field direction (and electron momentum distribution) breaks spatial
symmetry and acts as an optical axis. The $\perp$ (X, extraordinary) and $\parallel$
(O, ordinary) mode are defined as the electric field vector $\parallel$ or
$\perp$ to the plane containing the outgoing photon $\boldsymbol{k}_f$ and 
magnetic field $\boldsymbol{B}_{\rm loc}$ vectors,
respectively. There is an associated {\it{energy-dependent}} Doppler beaming
cone for electrons in the magnetosphere; the highest energy RICS
photons are sampled for electrons viewed head-on by an observer, corresponding to
lines of sight that are tangent to local field lines. Therefore, different viewing angles with respect to the
magnetic axis sample different electron populations and beaming
geometry. The upshot is spin modulation, i.e. (polarized) pulsations, if the spin
and magnetic moments are misaligned.

\begin{figure*}[t]
\begin{center}
\vspace{-5mm}
\includegraphics[angle=0,width=0.9\textwidth]{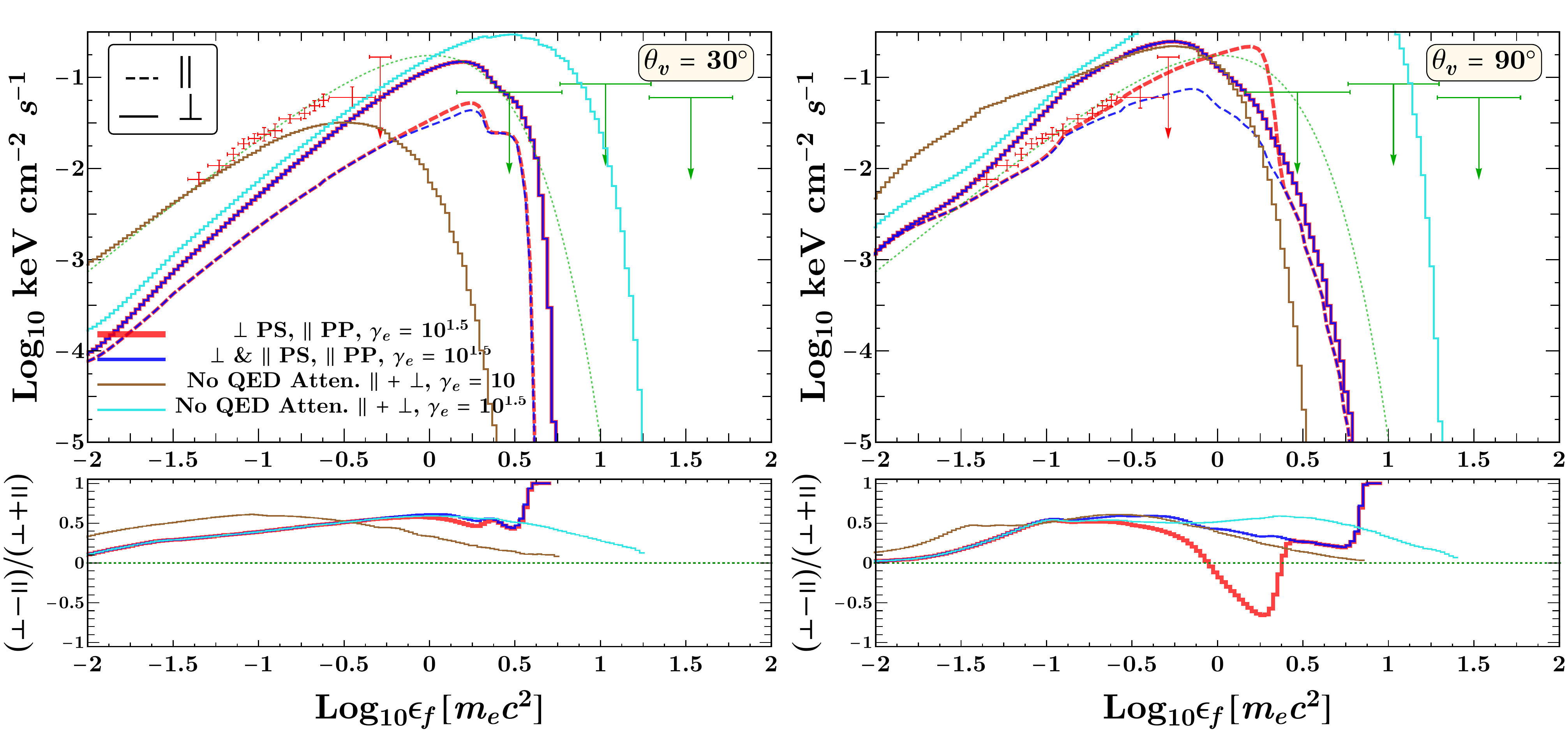}
\vspace{-3mm}
\caption{Spin-phase resolved model RICS spectra of a generic magnetar
  (at {\it{arbitrary normalization}}) overlaid on phase-averaged data
  for 4U 0412+61 along with a PL with exponential cutoff at $350$ keV in dotted green. The RICS emission is anticipated to be highly
  polarized and spin-phase dependent. The model emission is computed for surface photons of temperature $5\times 10^6$ K scattered by $\gamma_e = 10-10^{1.5}$ electrons uniformly populating field bundle from magnetic footpoint colatitudes $12 - 45^\circ$ for $B_p = 10 B_{\rm cr}$. {\bf{Left}}: Instantaneous observer impact
  angle (for a particular spin phase) of $\theta_v = 30^\circ$ with respect to the magnetic axis $\hat{\mu}$;
  {\bf{Right}}: $\theta_v = 90^\circ$. {\bf{Bottom panels}}: Signed
  polarization degree, highlighting the spectropolarimetric signatures of resonant Compton scattering attenuated by magnetic photon splitting (PS) and/or magnetic pair production (PP).} 
\label{ModelSpectra}
\vspace{-8mm}
\end{center}
\end{figure*}

\subparagraph{QED Propagation Effects:\,\,} Magnetar magnetospheres are
opaque to high energy photons, so that above the pair threshold around 1 MeV, pair creation strongly dominates the photon opacity. Dispersive influences of the magnetized quantum vacuum introduce birefringence, 
i.e. different refractive indices for the elliptical polarization eigenstates\cite[][]{2006RPPh...69.2631H}; 
dispersion is small for $\sim 1$~keV photons. Below
pair threshold, photon splitting is the dominant attenuation mechanism in a strong magnetic field; this
is a $3^{\rm rd}$ order QED process arising from vacuum polarization
(virtual pairs) radiating when interacting with the field. The rate of
splitting is a strong function of photon energy $\propto \epsilon^5
{\cal{B}}^6$ where ${\cal{B}}$ is the projection of the
local magnetic field $\boldsymbol{B}_{\rm loc}$ 
onto the direction of the photon momentum. In the
weakly dispersive limit, only $\perp$-mode photons may split due to
kinematic selection rules \citep{1971AnPhy..67..599A}. However, splitting of
both photon polarizations (modes) does not violate charge-parity (CP) symmetry; 
{\sl{it is still an open question if both modes may split in the strongly dispersive
    nonlinear regime of QED.}} If both polarizations are permitted to split,
then the {\it{shape}} of the spectral cutoff ought to follow a
super-exponential shape. 


In Fig.~\ref{ModelSpectra} we depict
selected RICS model spectra (Wadiasingh et al., in prep). For comparison, \integral\, data and COMPTEL bounds
for 4U 0142+61\cite{2008A&A...489..245D} are plotted along with a power law
with exponential cutoff at $350$~keV in dotted green. Hu et al. (2019, MNRAS submitted) provide a convenient parameterization of photon splitting and pair creation escape energies we use to
compute photon-trajectory-dependent opacities in our code for RICS
emission in Fig.~\ref{ModelSpectra}. As is typical
of scattering processes, the $\perp$ mode dominates for most energies
except near the unknown cutoff -- see the bottom panels.  Without
inclusion of QED opacities, the cutoff is kinematically attained and
exponential in character; this is illustrated in the
{\color{brown}{brown}} and {\color{cyan}{cyan}} polarization-summed
curves. In contrast, if the $\perp$ mode photon splitting and the $\parallel$ mode
pair creation attenuates the spectrum, a regime of very high polarization degree is exhibited in the cutoff as depicted
in the {\color{red}{red}} curves. Finally, if both $\perp$ and
$\parallel$ modes of splitting operate as represented by the {\color{blue}{blue}}
curves, then a depolarization effect in the cutoff is apparent, yet
with a cutoff that is no longer exponential but super-exponential.
Therefore, {\it spectropolarimetric diagnostics of the cutoff regime of
magnetars offer a powerful path to probing photon splitting}.  Also,
the sensitivity exhibited in Fig.~\ref{ModelSpectra} to $\theta_v$ indicates that 
phase-resolved spectropolarimetry will strongly constrain
the angle $\alpha$ between the rotational and magnetic axes of 
individual magnetars (Wadiasingh et al., in prep).







\section{Magnetar Soft Gamma-Ray Studies with Proposed Compton Technology}

\begin{wrapfigure}{r}{0.45\textwidth}
\vspace{-5mm}
\includegraphics[width=0.449\textwidth]{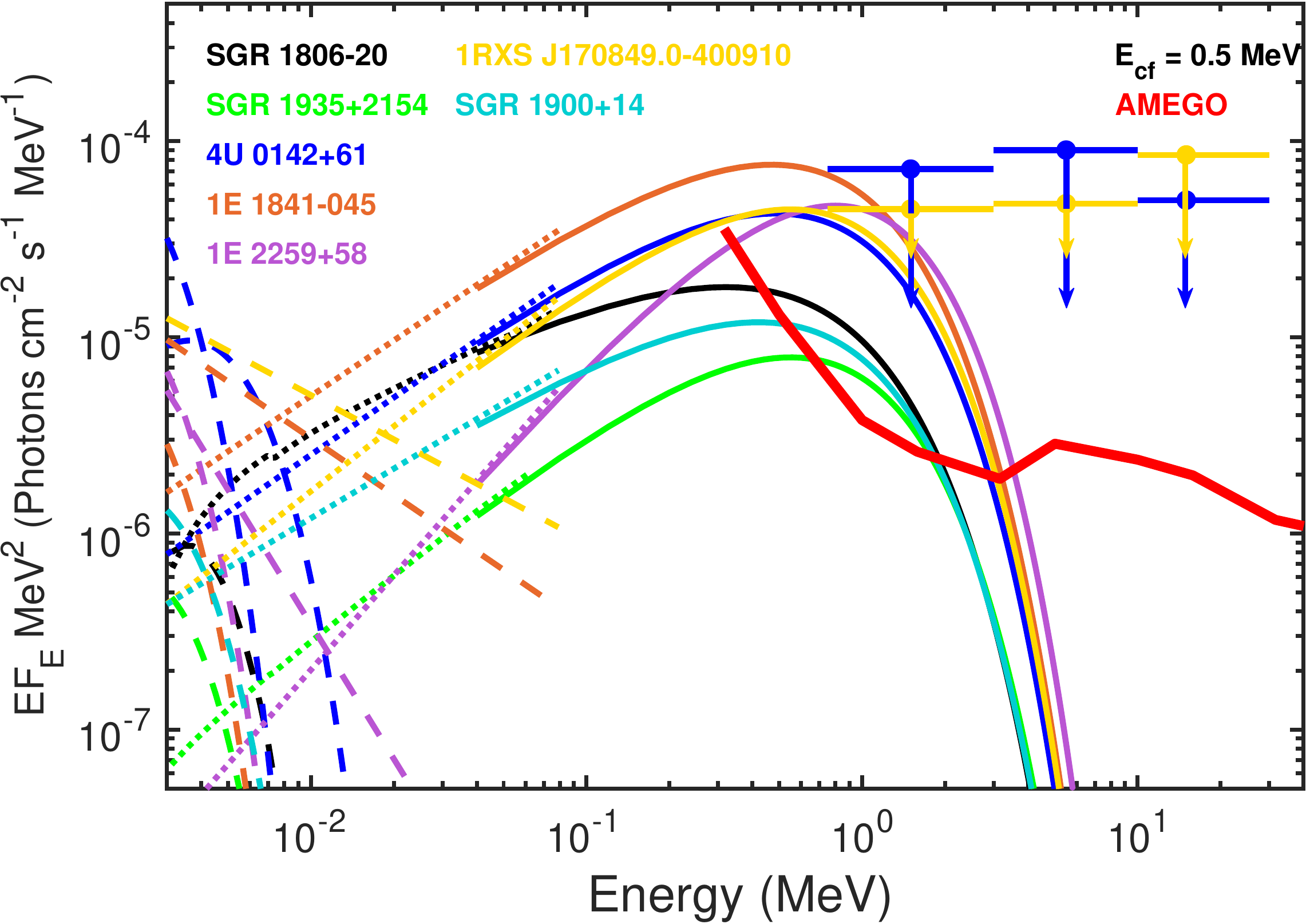}
\vspace{-5mm}
\caption{Extrapolation of \nustar \, and \integral \,  persistent phase-averaged PL
  spectra with a generic exponential cutoff at $0.5$~MeV for various
  magnetars.}
\vspace{-3mm}
 \label{detectability} 
 \end{wrapfigure}

\noindent There are currently 7 magnetars that exhibit a hard X-ray tail in
quiescence out of 8 observed with sensitive hard X-ray instruments such as
\suzaku\ and \nustar\ \citep{2017ApJS..231....8E}. These are depicted in
Fig.~\ref{detectability}. The energy scale covers $3\times10^{-3} - 40$ MeV, hence partially displays the soft X-ray quasi-thermal model (dashed-lines) and also hard power-law tails (dotted-lines). Our knowledge of these spectra extends
up to $\sim100 - 200$~keV (where it is {\textit{100\%~pulsed}}) beyond which our observational picture is completely missing. We extrapolated the observed phase-averaged hard tails of these magnetars and adopted an $0.5$~MeV exponential cutoff for all the sources. These mock soft $\gamma$-ray spectra are well below the $2\sigma$ CGRO-COMPTEL upper-limits for 4U~0142+61 and 1RXS~J170849.0$-$400910 depicted in blue and yellow, respectively. 

{\textit{Planned Compton telescope technology furnishes wide-field and polarization capabilities, enabling compelling time-domain and spectropolarimetric studies.}}
The red curve of
Fig.~\ref{detectability} denotes the 1-year sensitivity curve to the
{\sl{proposed probe-class mission AMEGO}}\cite{amego_2018_aa}(similar in capabilities to its proposed European kin
e-ASTROGAM\cite{2018JHEAp..19....1D}) and demonstrates that all seven
magnetars are well within the grasp of detectibility. Per our
theoretical predictions, the cutoff energy may be steeper with a
super-exponential shape. Similarly, simulation of a super-exponential
cutoff with index of 5 and a cutoff energy of 0.75 MeV results in the
detection of all seven magnetars. If some magnetar spectra extend beyond $1$ MeV, an advanced pair telescope such as AdEPT\cite{2014APh....59...18H,2018SPIE10699E..2MH} may reveal the presence of magnetic pair attenuation. Any instrument with {\sl AMEGO}-like large FoV
mission would also enable unprecedented spectral and temporal
variability studies of magnetars leading up to, during, and after outbursts, a crucial element to determining the excitation locales and triggering mechanism of these active states.


\begin{figure*}[t!]
\vspace{-5mm}
\begin{center}
\includegraphics[angle=0,width=0.45\textwidth]{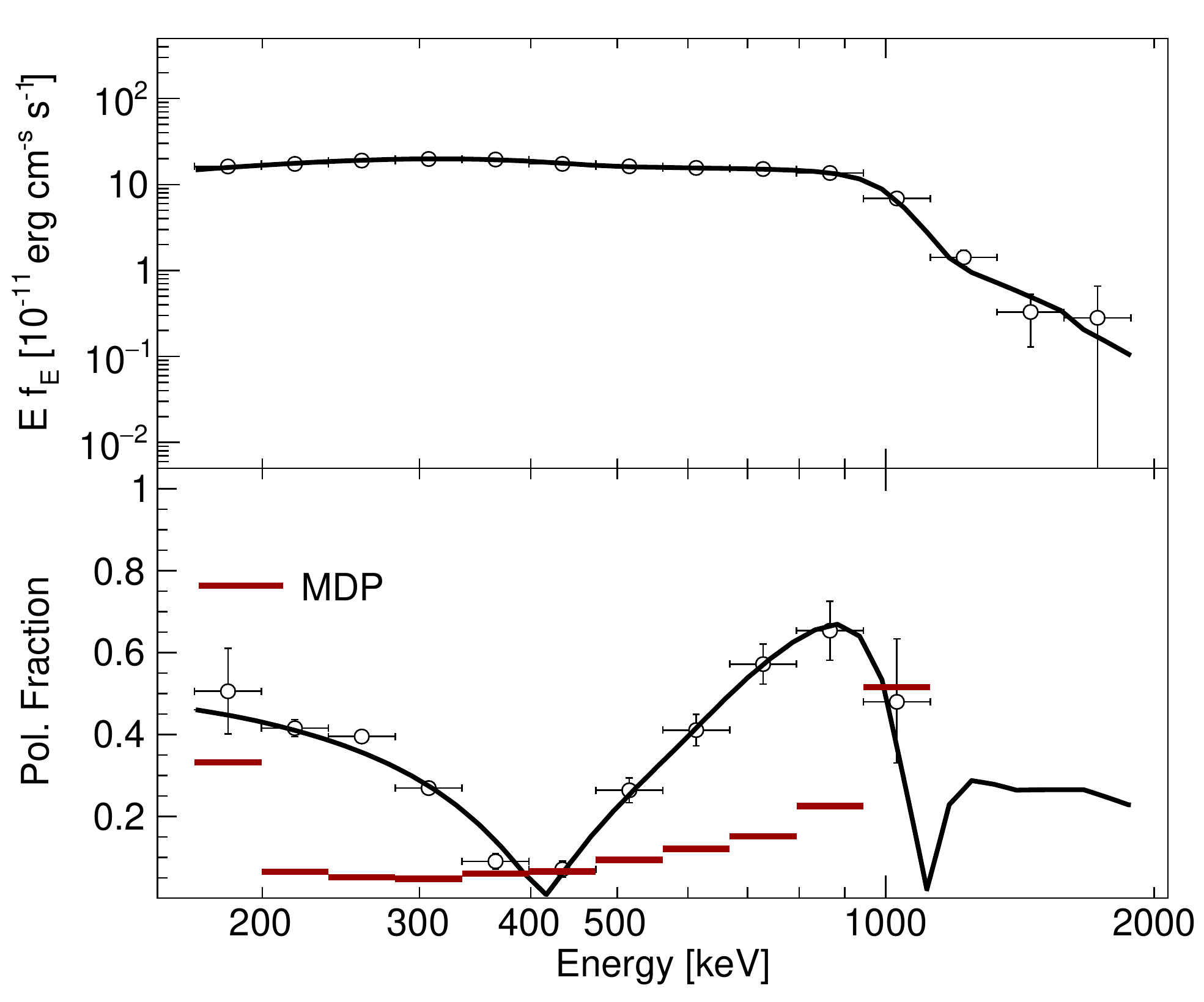}
\includegraphics[angle=0,width=0.45\textwidth]{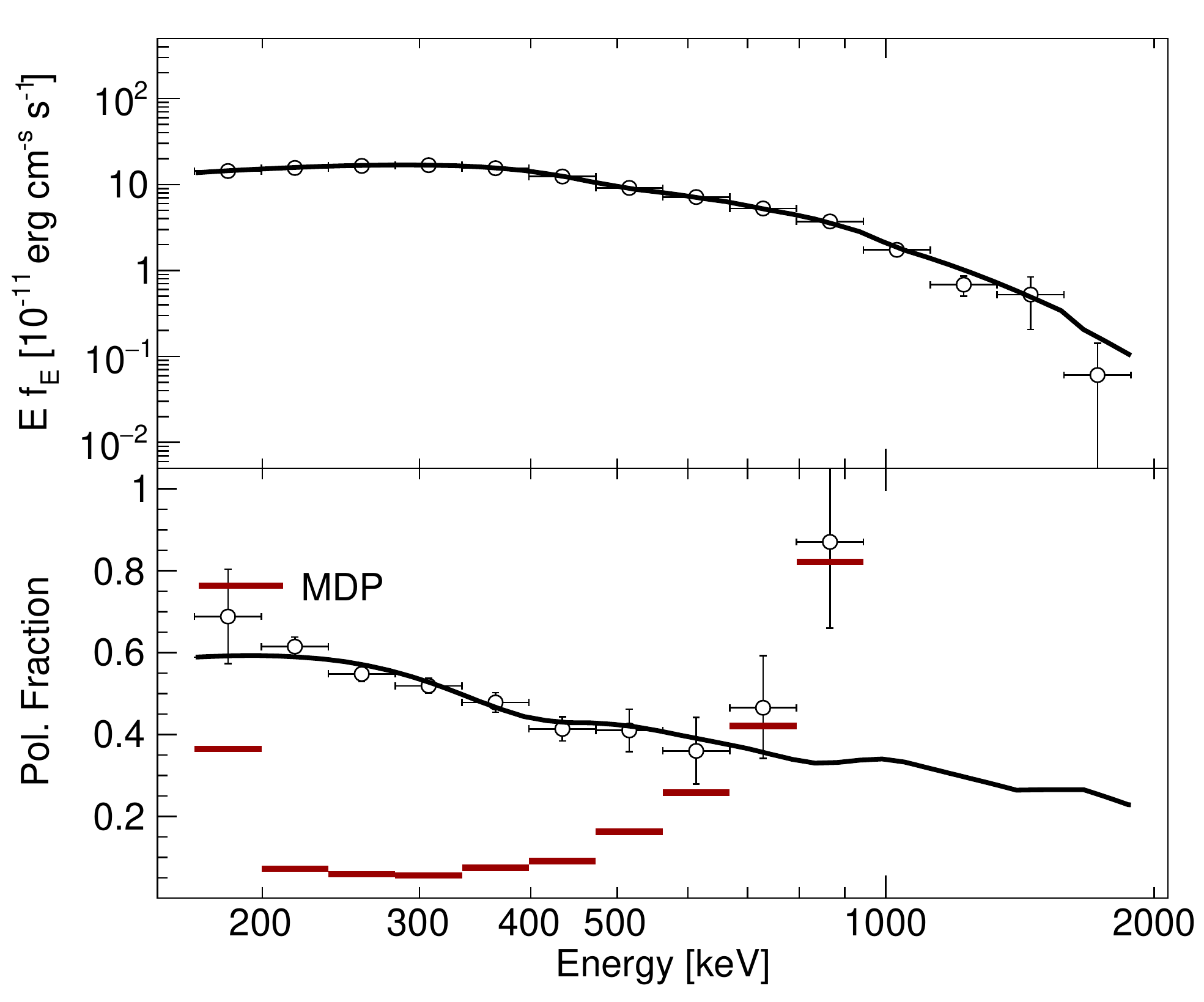}\\
\vspace{-3mm}
\caption{Simulations of model spectra and polarization (from the right
  panel of Fig.~\ref{ModelSpectra}) for $2$ yrs (146 days on-time) of
  observations with an AMEGO-like Compton telescope for a bright
  magnetar. ({\bf{Left panel}}) splitting of the $\perp$ (X) mode only. ({\bf{Right panel}}) splitting of
  both $\perp$ (X) and $\parallel$ (O) modes. Spectropolarimetry clearly offers a path for
  detecting and characterizing exotic QED splitting for the first time. MDP
  denotes the instrumental minimum detectable polarization level.}
\vspace{-9mm}
\label{simulation}
\end{center}
\end{figure*}

We also simulated, using GPST\citep{gpst_ref}, a 2-year
spectrum of the $\theta_v = 90^\circ$ models (tantamount to on-pulse phase selection) as displayed in Fig.~\ref{ModelSpectra} for an instrument with {\sl{AMEGO}}-like
spectropolarimetric capabilities (Fig.~\ref{simulation}). We assumed a
normalized flux of $2\times 10^{-11}$ erg s$^{-1}$ cm$^{-2}$ at 40
keV, consistent with the four brightest magnetars in
Fig.~\ref{detectability}. The left panel displays the case of photon splitting of the
$\perp$ mode only, while the right panel includes both splitting modes. In either
case, a detection of a high polarization degree would be a clear signature of the
resonant Compton model for the emission physics. Most importantly, the
cutoff shape and polarization differentiation between the two cases are
unmistakable. {\sl{A detection of a polarization signature similar to
    the one depicted in the left panel is a ``smoking-gun'' evidence of
    photon splitting, and would be a spectacular confirmation of QED in the strong
    field domain.}} 

\section{Summary}

\noindent
The case is presented that a new, sensitive Compton telescope with
polarimetric capacity will move our understanding of magnetars and the
physics of their magnetospheres forward in a watershed fashion. For
the first time, the energy budget apportioned to their non-thermal
signals will be measured with good precision.  Pulse phase-resolved
spectroscopy and polarimetry will constrain the locales possible for
the origin of the hard X-ray emission.  Spectropolarimetry will also enable
the estimation of the inclination angle $\alpha$ between the magnetic
and rotation axes, a key magnetar parameter. Moreover, these
observational capabilities will determine whether or not the exotic
QED process of photon splitting in strong magnetic fields is operating in Nature.  Magnetars will thus serve as a cosmic laboratory that opens windows into the
physical Universe that are not presently afforded by terrestrial
experiments.  The technology that can bring about these gains for
magnetars is on the near horizon, and can be applied to a multitude of
other astronomical source classes with similar prospects for new and
advanced insights. 

\vspace{-5mm}

\newpage
\noindent This work has made use of the NASA Astrophysics Data System.
\setcounter{page}{1}
\noindent
\bibliographystyle{JHEP}
\bibliography{references}

\providecommand{\href}[2]{#2}\begingroup\raggedright\begin{thebibliography}{10}

\bibitem{2014ApJS..212....6O}
S.~A. {Olausen} and V.~M. {Kaspi}, {\it {The McGill Magnetar Catalog}},  {\em
  \apjs} {\bf 212} (May, 2014) 6, [\href{http://arxiv.org/abs/1309.4167}{{\tt
  arXiv:1309.4167}}].

\bibitem{zelati18MNRAS:magOut}
F.~{Coti Zelati}, N.~{Rea}, J.~A. {Pons}, S.~{Campana}, and P.~{Esposito}, {\it
  {Systematic study of magnetar outbursts}},  {\em \mnras} {\bf 474} (Feb.,
  2018) 961--1017, [\href{http://arxiv.org/abs/1710.04671}{{\tt
  arXiv:1710.04671}}].

\bibitem{2017ApJS..231....8E}
T.~{Enoto}, S.~{Shibata}, T.~{Kitaguchi}, Y.~{Suwa}, T.~{Uchide},
  H.~{Nishioka}, S.~{Kisaka}, T.~{Nakano}, H.~{Murakami}, and K.~{Makishima},
  {\it {Magnetar Broadband X-Ray Spectra Correlated with Magnetic Fields:
  Suzaku Archive of SGRs and AXPs Combined with NuSTAR, Swift, and RXTE}},
  {\em \apjs} {\bf 231} (July, 2017) 8,
  [\href{http://arxiv.org/abs/1704.07018}{{\tt arXiv:1704.07018}}].

\bibitem{woods04ApJ:1E2259}
P.~M. {Woods}, V.~M. {Kaspi}, C.~{Thompson}, F.~P. {Gavriil}, H.~L. {Marshall},
  D.~{Chakrabarty}, K.~{Flanagan}, J.~{Heyl}, and L.~{Hernquist}, {\it {Changes
  in the X-Ray Emission from the Magnetar Candidate 1E 2259+586 during Its 2002
  Outburst}},  {\em \apj} {\bf 605} (Apr., 2004) 378--399,
  [\href{http://arxiv.org/abs/astro-ph/0310575}{{\tt astro-ph/0310575}}].

\bibitem{israel07ApJ:1647}
G.~L. {Israel}, S.~{Campana}, S.~{Dall'Osso}, M.~P. {Muno}, J.~{Cummings},
  R.~{Perna}, and L.~{Stella}, {\it {The Post-Burst Awakening of the Anomalous
  X-Ray Pulsar in Westerlund 1}},  {\em \apj} {\bf 664} (July, 2007) 448--457,
  [\href{http://arxiv.org/abs/astro-ph/0703684}{{\tt astro-ph/0703684}}].

\bibitem{archibald13Natur}
R.~F. {Archibald}, V.~M. {Kaspi}, C.-Y. {Ng}, K.~N. {Gourgouliatos},
  D.~{Tsang}, P.~{Scholz}, A.~P. {Beardmore}, N.~{Gehrels}, and J.~A. {Kennea},
  {\it {An anti-glitch in a magnetar}},  {\em \nat} {\bf 497} (May, 2013)
  591--593, [\href{http://arxiv.org/abs/1305.6894}{{\tt arXiv:1305.6894}}].

\bibitem{rea13ApJ:0418}
N.~{Rea}, G.~L. {Israel}, J.~A. {Pons}, R.~{Turolla}, D.~{Vigan{\`o}},
  S.~{Zane}, P.~{Esposito}, R.~{Perna}, A.~{Papitto}, G.~{Terreran},
  A.~{Tiengo}, D.~{Salvetti}, J.~M. {Girart}, A.~{Palau}, A.~{Possenti},
  M.~{Burgay}, E.~{G{\"o}{\u g}{\"u}{\c s}}, G.~A. {Caliandro},
  C.~{Kouveliotou}, D.~{G{\"o}tz}, R.~P. {Mignani}, E.~{Ratti}, and
  L.~{Stella}, {\it {The Outburst Decay of the Low Magnetic Field Magnetar SGR
  0418+5729}},  {\em \apj} {\bf 770} (June, 2013) 65,
  [\href{http://arxiv.org/abs/1303.5579}{{\tt arXiv:1303.5579}}].

\bibitem{kaspi14ApJ:1745}
V.~M. {Kaspi}, R.~F. {Archibald}, V.~{Bhalerao}, F.~{Dufour}, E.~V. {Gotthelf},
  H.~{An}, M.~{Bachetti}, A.~M. {Beloborodov}, S.~E. {Boggs}, F.~E.
  {Christensen}, W.~W. {Craig}, B.~W. {Grefenstette}, C.~J. {Hailey}, F.~A.
  {Harrison}, J.~A. {Kennea}, C.~{Kouveliotou}, K.~K. {Madsen}, K.~{Mori},
  C.~B. {Markwardt}, D.~{Stern}, J.~K. {Vogel}, and W.~W. {Zhang}, {\it {Timing
  and Flux Evolution of the Galactic Center Magnetar SGR J1745-2900}},  {\em
  \apj} {\bf 786} (May, 2014) 84, [\href{http://arxiv.org/abs/1403.5344}{{\tt
  arXiv:1403.5344}}].

\bibitem{zelati15MNRAS:1745}
F.~{Coti Zelati}, N.~{Rea}, A.~{Papitto}, D.~{Vigan{\`o}}, J.~A. {Pons},
  R.~{Turolla}, P.~{Esposito}, D.~{Haggard}, F.~K. {Baganoff}, G.~{Ponti},
  G.~L. {Israel}, S.~{Campana}, D.~F. {Torres}, A.~{Tiengo}, S.~{Mereghetti},
  R.~{Perna}, S.~{Zane}, R.~P. {Mignani}, A.~{Possenti}, and L.~{Stella}, {\it
  {The X-ray outburst of the Galactic Centre magnetar SGR J1745-2900 during the
  first 1.5 year}},  {\em \mnras} {\bf 449} (May, 2015) 2685--2699,
  [\href{http://arxiv.org/abs/1503.01307}{{\tt arXiv:1503.01307}}].

\bibitem{scholz14ApJ:1822}
P.~{Scholz}, V.~M. {Kaspi}, and A.~{Cumming}, {\it {The Long-term Post-outburst
  Spin Down and Flux Relaxation of Magnetar Swift J1822.3-1606}},  {\em \apj}
  {\bf 786} (May, 2014) 62, [\href{http://arxiv.org/abs/1401.6965}{{\tt
  arXiv:1401.6965}}].

\bibitem{alford16ApJ:1810}
J.~A.~J. {Alford} and J.~P. {Halpern}, {\it {Evolution of the X-Ray Properties
  of the Transient Magnetar XTE J1810-197}},  {\em \apj} {\bf 818} (Feb., 2016)
  122, [\href{http://arxiv.org/abs/1601.00757}{{\tt arXiv:1601.00757}}].

\bibitem{younes17:1935}
G.~{Younes}, C.~{Kouveliotou}, A.~{Jaodand}, M.~G. {Baring}, A.~J. {van der
  Horst}, A.~K. {Harding}, J.~W.~T. {Hessels}, N.~{Gehrels}, R.~{Gill},
  D.~{Huppenkothen}, J.~{Granot}, E.~{G{\"o}{\u g}{\"u}{\c s}}, and L.~{Lin},
  {\it {X-Ray and Radio Observations of the Magnetar SGR J1935+2154 during Its
  2014, 2015, and 2016 Outbursts}},  {\em \apj} {\bf 847} (Oct., 2017) 85,
  [\href{http://arxiv.org/abs/1702.04370}{{\tt arXiv:1702.04370}}].

\bibitem{2017ApJ...851...17Y}
G.~{Younes}, M.~G. {Baring}, C.~{Kouveliotou}, A.~{Harding}, S.~{Donovan},
  E.~{G{\"o}{\u g}{\"u}{\c s}}, V.~{Kaspi}, and J.~{Granot}, {\it {The Sleeping
  Monster: NuSTAR Observations of SGR 1806-20, 11 Years After the Giant
  Flare}},  {\em \apj} {\bf 851} (Dec., 2017) 17,
  [\href{http://arxiv.org/abs/1711.00034}{{\tt arXiv:1711.00034}}].

\bibitem{mereghetti08AARv:magentars}
S.~{Mereghetti}, {\it {The strongest cosmic magnets: soft gamma-ray repeaters
  and anomalous X-ray pulsars}},  {\em \aapr} {\bf 15} (July, 2008) 225--287,
  [\href{http://arxiv.org/abs/0804.0250}{{\tt arXiv:0804.0250}}].

\bibitem{rea11:outburst}
N.~{Rea} and P.~{Esposito}, {\it {Magnetar outbursts: an observational
  review}},  in {\em High-Energy Emission from Pulsars and their Systems}
  (D.~F. {Torres} and N.~{Rea}, eds.), p.~247, 2011.
\newblock \href{http://arxiv.org/abs/1101.4472}{{\tt arXiv:1101.4472}}.

\bibitem{turolla15:mag}
R.~{Turolla}, S.~{Zane}, and A.~L. {Watts}, {\it {Magnetars: the physics behind
  observations. A review}},  {\em Reports on Progress in Physics} {\bf 78}
  (Nov., 2015) 116901, [\href{http://arxiv.org/abs/1507.02924}{{\tt
  arXiv:1507.02924}}].

\bibitem{mereghetti15:mag}
S.~{Mereghetti}, J.~A. {Pons}, and A.~{Melatos}, {\it {Magnetars: Properties,
  Origin and Evolution}},  {\em \ssr} {\bf 191} (Oct., 2015) 315--338,
  [\href{http://arxiv.org/abs/1503.06313}{{\tt arXiv:1503.06313}}].

\bibitem{kaspi17:magnetars}
V.~M. {Kaspi} and A.~M. {Beloborodov}, {\it {Magnetars}},  {\em \araa} {\bf 55}
  (Aug., 2017) 261--301, [\href{http://arxiv.org/abs/1703.00068}{{\tt
  arXiv:1703.00068}}].

\bibitem{esposito18:rev}
P.~{Esposito}, N.~{Rea}, and G.~L. {Israel}, {\it {Magnetars: a short review
  and some sparse considerations}},  {\em arXiv e-prints} (Mar, 2018)
  arXiv:1803.05716, [\href{http://arxiv.org/abs/1803.05716}{{\tt
  arXiv:1803.05716}}].

\bibitem{kasen10ApJ:SNMag}
D.~{Kasen} and L.~{Bildsten}, {\it {Supernova Light Curves Powered by Young
  Magnetars}},  {\em \apj} {\bf 717} (July, 2010) 245--249,
  [\href{http://arxiv.org/abs/0911.0680}{{\tt arXiv:0911.0680}}].

\bibitem{inserra13ApJ:SNMag}
C.~{Inserra}, S.~J. {Smartt}, A.~{Jerkstrand}, S.~{Valenti}, M.~{Fraser},
  D.~{Wright}, K.~{Smith}, T.-W. {Chen}, R.~{Kotak}, A.~{Pastorello},
  M.~{Nicholl}, F.~{Bresolin}, R.~P. {Kudritzki}, S.~{Benetti}, M.~T.
  {Botticella}, W.~S. {Burgett}, K.~C. {Chambers}, M.~{Ergon}, H.~{Flewelling},
  J.~P.~U. {Fynbo}, S.~{Geier}, K.~W. {Hodapp}, D.~A. {Howell}, M.~{Huber},
  N.~{Kaiser}, G.~{Leloudas}, L.~{Magill}, E.~A. {Magnier}, M.~G. {McCrum},
  N.~{Metcalfe}, P.~A. {Price}, A.~{Rest}, J.~{Sollerman}, W.~{Sweeney},
  F.~{Taddia}, S.~{Taubenberger}, J.~L. {Tonry}, R.~J. {Wainscoat},
  C.~{Waters}, and D.~{Young}, {\it {Super-luminous Type Ic Supernovae:
  Catching a Magnetar by the Tail}},  {\em \apj} {\bf 770} (June, 2013) 128,
  [\href{http://arxiv.org/abs/1304.3320}{{\tt arXiv:1304.3320}}].

\bibitem{wheeler00ApJ}
J.~C. {Wheeler}, I.~{Yi}, P.~{H{\"o}flich}, and L.~{Wang}, {\it {Asymmetric
  Supernovae, Pulsars, Magnetars, and Gamma-Ray Bursts}},  {\em \apj} {\bf 537}
  (July, 2000) 810--823, [\href{http://arxiv.org/abs/astro-ph/9909293}{{\tt
  astro-ph/9909293}}].

\bibitem{rowlinson13MNRAS}
A.~{Rowlinson}, P.~T. {O'Brien}, B.~D. {Metzger}, N.~R. {Tanvir}, and A.~J.
  {Levan}, {\it {Signatures of magnetar central engines in short GRB light
  curves}},  {\em \mnras} {\bf 430} (Apr., 2013) 1061--1087,
  [\href{http://arxiv.org/abs/1301.0629}{{\tt arXiv:1301.0629}}].

\bibitem{2014Natur.514..202B}
M.~{Bachetti}, F.~A. {Harrison}, D.~J. {Walton}, B.~W. {Grefenstette},
  D.~{Chakrabarty}, F.~{F{\"u}rst}, D.~{Barret}, A.~{Beloborodov}, S.~E.
  {Boggs}, F.~E. {Christensen}, W.~W. {Craig}, A.~C. {Fabian}, C.~J. {Hailey},
  A.~{Hornschemeier}, V.~{Kaspi}, S.~R. {Kulkarni}, T.~{Maccarone}, J.~M.
  {Miller}, V.~{Rana}, D.~{Stern}, S.~P. {Tendulkar}, J.~{Tomsick}, N.~A.
  {Webb}, and W.~W. {Zhang}, {\it {An ultraluminous X-ray source powered by an
  accreting neutron star}},  {\em \nat} {\bf 514} (Oct., 2014) 202--204,
  [\href{http://arxiv.org/abs/1410.3590}{{\tt arXiv:1410.3590}}].

\bibitem{israel17Sci:ULXmag}
G.~L. {Israel}, A.~{Belfiore}, L.~{Stella}, P.~{Esposito}, P.~{Casella}, A.~{De
  Luca}, M.~{Marelli}, A.~{Papitto}, M.~{Perri}, S.~{Puccetti}, G.~A.~R.
  {Castillo}, D.~{Salvetti}, A.~{Tiengo}, L.~{Zampieri}, D.~{D'Agostino},
  J.~{Greiner}, F.~{Haberl}, G.~{Novara}, R.~{Salvaterra}, R.~{Turolla},
  M.~{Watson}, J.~{Wilms}, and A.~{Wolter}, {\it {An accreting pulsar with
  extreme properties drives an ultraluminous x-ray source in NGC 5907}},  {\em
  Science} {\bf 355} (Feb., 2017) 817--819,
  [\href{http://arxiv.org/abs/1609.07375}{{\tt arXiv:1609.07375}}].

\bibitem{ekcsi15MNRAS:ulxmag}
K.~Y. {Ek{\c s}i}, {\.I}.~C. {Anda{\c c}}, S.~{{\c C}{\i}k{\i}nto{\u g}lu},
  A.~A. {Gen{\c c}ali}, C.~{G{\"u}ng{\"o}r}, and F.~{{\"O}ztekin}, {\it {The
  ultraluminous X-ray source NuSTAR J095551+6940.8: a magnetar in a high-mass
  X-ray binary}},  {\em \mnras} {\bf 448} (Mar., 2015) L40--L42,
  [\href{http://arxiv.org/abs/1410.5205}{{\tt arXiv:1410.5205}}].

\bibitem{dallosso15MNRAS}
S.~{Dall'Osso}, R.~{Perna}, and L.~{Stella}, {\it {NuSTAR J095551+6940.8: a
  highly magnetized neutron star with super-Eddington mass accretion}},  {\em
  \mnras} {\bf 449} (May, 2015) 2144--2150,
  [\href{http://arxiv.org/abs/1412.1823}{{\tt arXiv:1412.1823}}].

\bibitem{2014MNRAS.442L...9L}
Y.~{Lyubarsky}, {\it {A model for fast extragalactic radio bursts}},  {\em
  \mnras} {\bf 442} (July, 2014) L9--L13,
  [\href{http://arxiv.org/abs/1401.6674}{{\tt arXiv:1401.6674}}].

\bibitem{masui15Natur}
K.~{Masui}, H.-H. {Lin}, J.~{Sievers}, C.~J. {Anderson}, T.-C. {Chang},
  X.~{Chen}, A.~{Ganguly}, M.~{Jarvis}, C.-Y. {Kuo}, Y.-C. {Li}, Y.-W. {Liao},
  M.~{McLaughlin}, U.-L. {Pen}, J.~B. {Peterson}, A.~{Roman}, P.~T. {Timbie},
  T.~{Voytek}, and J.~K. {Yadav}, {\it {Dense magnetized plasma associated with
  a fast radio burst}},  {\em \nat} {\bf 528} (Dec., 2015) 523--525,
  [\href{http://arxiv.org/abs/1512.00529}{{\tt arXiv:1512.00529}}].

\bibitem{2016ApJ...826..226K}
J.~I. {Katz}, {\it {How Soft Gamma Repeaters Might Make Fast Radio Bursts}},
  {\em \apj} {\bf 826} (Aug., 2016) 226,
  [\href{http://arxiv.org/abs/1512.04503}{{\tt arXiv:1512.04503}}].

\bibitem{metzger17ApJ}
B.~D. {Metzger}, E.~{Berger}, and B.~{Margalit}, {\it {Millisecond Magnetar
  Birth Connects FRB 121102 to Superluminous Supernovae and Long-duration
  Gamma-Ray Bursts}},  {\em \apj} {\bf 841} (May, 2017) 14,
  [\href{http://arxiv.org/abs/1701.02370}{{\tt arXiv:1701.02370}}].

\bibitem{2017ApJ...843L..26B}
A.~M. {Beloborodov}, {\it {A Flaring Magnetar in FRB 121102?}},  {\em \apjl}
  {\bf 843} (July, 2017) L26, [\href{http://arxiv.org/abs/1702.08644}{{\tt
  arXiv:1702.08644}}].

\bibitem{horst10ApJ:0418}
A.~J. {van der Horst}, V.~{Connaughton}, C.~{Kouveliotou}, E.~{G{\"o}{\v
  g}{\"u}{\c s}}, Y.~{Kaneko}, S.~{Wachter}, M.~S. {Briggs}, J.~{Granot},
  E.~{Ramirez-Ruiz}, P.~M. {Woods}, R.~L. {Aptekar}, S.~D. {Barthelmy}, J.~R.
  {Cummings}, M.~H. {Finger}, D.~D. {Frederiks}, N.~{Gehrels}, C.~R. {Gelino},
  D.~M. {Gelino}, S.~{Golenetskii}, K.~{Hurley}, H.~A. {Krimm}, E.~P. {Mazets},
  J.~E. {McEnery}, C.~A. {Meegan}, P.~P. {Oleynik}, D.~M. {Palmer}, V.~D.
  {Pal'shin}, A.~{Pe'er}, D.~{Svinkin}, M.~V. {Ulanov}, M.~{van der Klis},
  A.~{von Kienlin}, A.~L. {Watts}, and C.~A. {Wilson-Hodge}, {\it {Discovery of
  a New Soft Gamma Repeater: SGR J0418 + 5729}},  {\em \apjl} {\bf 711} (Mar.,
  2010) L1--L6, [\href{http://arxiv.org/abs/0911.5544}{{\tt arXiv:0911.5544}}].

\bibitem{kennea13ApJ}
J.~A. {Kennea}, D.~N. {Burrows}, C.~{Kouveliotou}, D.~M. {Palmer},
  E.~{G{\"o}{\u g}{\"u}{\c s}}, Y.~{Kaneko}, P.~A. {Evans}, N.~{Degenaar},
  M.~T. {Reynolds}, J.~M. {Miller}, R.~{Wijnands}, K.~{Mori}, and N.~{Gehrels},
  {\it {Swift Discovery of a New Soft Gamma Repeater, SGR J1745-29, near
  Sagittarius A*}},  {\em \apjl} {\bf 770} (June, 2013) L24,
  [\href{http://arxiv.org/abs/1305.2128}{{\tt arXiv:1305.2128}}].

\bibitem{woods07ApJ:1906}
P.~M. {Woods}, C.~{Kouveliotou}, M.~H. {Finger}, E.~{G{\"o}{\v g}{\"u}{\c s}},
  C.~A. {Wilson}, S.~K. {Patel}, K.~{Hurley}, and J.~H. {Swank}, {\it {The
  Prelude to and Aftermath of the Giant Flare of 2004 December 27: Persistent
  and Pulsed X-Ray Properties of SGR 1806-20 from 1993 to 2005}},  {\em \apj}
  {\bf 654} (Jan., 2007) 470--486,
  [\href{http://arxiv.org/abs/astro-ph/0602402}{{\tt astro-ph/0602402}}].

\bibitem{cotizelati18MNRAS}
F.~{Coti Zelati}, N.~{Rea}, J.~A. {Pons}, S.~{Campana}, and P.~{Esposito}, {\it
  {Systematic study of magnetar outbursts}},  {\em \mnras} {\bf 474} (Feb.,
  2018) 961--1017, [\href{http://arxiv.org/abs/1710.04671}{{\tt
  arXiv:1710.04671}}].

\bibitem{2004ApJ...613.1173K}
L.~{Kuiper}, W.~{Hermsen}, and M.~{Mendez}, {\it {Discovery of Hard Nonthermal
  Pulsed X-Ray Emission from the Anomalous X-Ray Pulsar 1E 1841-045}},  {\em
  \apj} {\bf 613} (Oct., 2004) 1173--1178,
  [\href{http://arxiv.org/abs/astro-ph/0404582}{{\tt astro-ph/0404582}}].

\bibitem{2006ApJ...645..556K}
L.~{Kuiper}, W.~{Hermsen}, P.~R. {den Hartog}, and W.~{Collmar}, {\it
  {Discovery of Luminous Pulsed Hard X-Ray Emission from Anomalous X-Ray
  Pulsars 1RXS J1708-4009, 4U 0142+61, and 1E 2259+586 by INTEGRAL and RXTE}},
  {\em \apj} {\bf 645} (July, 2006) 556--575,
  [\href{http://arxiv.org/abs/astro-ph/0603467}{{\tt astro-ph/0603467}}].

\bibitem{2008A&A...489..245D}
P.~R. {den Hartog}, L.~{Kuiper}, W.~{Hermsen}, V.~M. {Kaspi}, R.~{Dib},
  J.~{Kn{\"o}dlseder}, and F.~P. {Gavriil}, {\it {Detailed high-energy
  characteristics of AXP 4U 0142+61. Multi-year observations with INTEGRAL,
  RXTE, XMM-Newton, and ASCA}},  {\em \aap} {\bf 489} (Oct., 2008) 245--261,
  [\href{http://arxiv.org/abs/0804.1640}{{\tt arXiv:0804.1640}}].

\bibitem{2008A&A...489..263D}
P.~R. {den Hartog}, L.~{Kuiper}, and W.~{Hermsen}, {\it {Detailed high-energy
  characteristics of AXP 1RXS J170849-400910. Probing the magnetosphere using
  INTEGRAL, RXTE, and XMM-Newton}},  {\em \aap} {\bf 489} (Oct., 2008)
  263--279, [\href{http://arxiv.org/abs/0804.1641}{{\tt arXiv:0804.1641}}].

\bibitem{2007ApJ...657..967B}
A.~M. {Beloborodov} and C.~{Thompson}, {\it {Corona of Magnetars}},  {\em \apj}
  {\bf 657} (Mar., 2007) 967--993,
  [\href{http://arxiv.org/abs/astro-ph/0602417}{{\tt astro-ph/0602417}}].

\bibitem{2013ApJ...777..114B}
A.~M. {Beloborodov}, {\it {Electron-Positron Flows around Magnetars}},  {\em
  \apj} {\bf 777} (Nov., 2013) 114, [\href{http://arxiv.org/abs/1209.4063}{{\tt
  arXiv:1209.4063}}].

\bibitem{2007Ap&SS.308..109B}
M.~G. {Baring} and A.~K. {Harding}, {\it {Resonant Compton upscattering in
  anomalous X-ray pulsars}},  {\em \apss} {\bf 308} (Apr., 2007) 109--118,
  [\href{http://arxiv.org/abs/astro-ph/0610382}{{\tt astro-ph/0610382}}].

\bibitem{2007ApJ...660..615F}
R.~{Fern{\'a}ndez} and C.~{Thompson}, {\it {Resonant Cyclotron Scattering in
  Three Dimensions and the Quiescent Nonthermal X-ray Emission of Magnetars}},
  {\em \apj} {\bf 660} (May, 2007) 615--640,
  [\href{http://arxiv.org/abs/astro-ph/0608281}{{\tt astro-ph/0608281}}].

\bibitem{2008MNRAS.389..989N}
L.~{Nobili}, R.~{Turolla}, and S.~{Zane}, {\it {X-ray spectra from magnetar
  candidates - II. Resonant cross-sections for electron-photon scattering in
  the relativistic regime}},  {\em \mnras} {\bf 389} (Sept., 2008) 989--1000,
  [\href{http://arxiv.org/abs/0806.3714}{{\tt arXiv:0806.3714}}].

\bibitem{2011AdSpR..47.1298Z}
S.~{Zane}, R.~{Turolla}, L.~{Nobili}, and N.~{Rea}, {\it {Modeling the
  broadband persistent emission of magnetars}},  {\em Advances in Space
  Research} {\bf 47} (Apr., 2011) 1298--1304,
  [\href{http://arxiv.org/abs/1008.1537}{{\tt arXiv:1008.1537}}].

\bibitem{2011ApJ...733...61B}
M.~G. {Baring}, Z.~{Wadiasingh}, and P.~L. {Gonthier}, {\it {Cooling Rates for
  Relativistic Electrons Undergoing Compton Scattering in Strong Magnetic
  Fields}},  {\em \apj} {\bf 733} (May, 2011) 61,
  [\href{http://arxiv.org/abs/1103.3356}{{\tt arXiv:1103.3356}}].

\bibitem{2013ApJ...762...13B}
A.~M. {Beloborodov}, {\it {On the Mechanism of Hard X-Ray Emission from
  Magnetars}},  {\em \apj} {\bf 762} (Jan., 2013) 13,
  [\href{http://arxiv.org/abs/1201.0664}{{\tt arXiv:1201.0664}}].

\bibitem{2014ApJ...786L...1H}
R.~{Hasco{\"e}t}, A.~M. {Beloborodov}, and P.~R. {den Hartog}, {\it
  {Phase-resolved X-Ray Spectra of Magnetars and the Coronal Outflow Model}},
  {\em \apjl} {\bf 786} (May, 2014) L1,
  [\href{http://arxiv.org/abs/1401.3406}{{\tt arXiv:1401.3406}}].

\bibitem{2018ApJ...854...98W}
Z.~{Wadiasingh}, M.~G. {Baring}, P.~L. {Gonthier}, and A.~K. {Harding}, {\it
  {Resonant Inverse Compton Scattering Spectra from Highly Magnetized Neutron
  Stars}},  {\em \apj} {\bf 854} (Feb., 2018) 98,
  [\href{http://arxiv.org/abs/1712.09643}{{\tt arXiv:1712.09643}}].

\bibitem{1997ApJ...476..246H}
A.~K. {Harding}, M.~G. {Baring}, and P.~L. {Gonthier}, {\it {Photon-Splitting
  Cascades in Gamma-Ray Pulsars and the Spectrum of PSR 1509-58}},  {\em \apj}
  {\bf 476} (Feb., 1997) 246--260,
  [\href{http://arxiv.org/abs/astro-ph/9609167}{{\tt astro-ph/9609167}}].

\bibitem{2001ApJ...547..929B}
M.~G. {Baring} and A.~K. {Harding}, {\it {Photon Splitting and Pair Creation in
  Highly Magnetized Pulsars}},  {\em \apj} {\bf 547} (Feb., 2001) 929--948,
  [\href{http://arxiv.org/abs/astro-ph/0010400}{{\tt astro-ph/0010400}}].

\bibitem{2006RPPh...69.2631H}
A.~K. {Harding} and D.~{Lai}, {\it {Physics of strongly magnetized neutron
  stars}},  {\em Reports on Progress in Physics} {\bf 69} (Sept., 2006)
  2631--2708, [\href{http://arxiv.org/abs/astro-ph/0606674}{{\tt
  astro-ph/0606674}}].

\bibitem{2000ApJ...540..907G}
P.~L. {Gonthier}, A.~K. {Harding}, M.~G. {Baring}, R.~M. {Costello}, and C.~L.
  {Mercer}, {\it {Compton Scattering in Ultrastrong Magnetic Fields: Numerical
  and Analytical Behavior in the Relativistic Regime}},  {\em \apj} {\bf 540}
  (Sept., 2000) 907--922, [\href{http://arxiv.org/abs/astro-ph/0005072}{{\tt
  astro-ph/0005072}}].

\bibitem{2005ApJ...630..430B}
M.~G. {Baring}, P.~L. {Gonthier}, and A.~K. {Harding}, {\it {Spin-dependent
  Cyclotron Decay Rates in Strong Magnetic Fields}},  {\em \apj} {\bf 630}
  (Sept., 2005) 430--440, [\href{http://arxiv.org/abs/astro-ph/0505327}{{\tt
  astro-ph/0505327}}].

\bibitem{2014PhRvD..90d3014G}
P.~L. {Gonthier}, M.~G. {Baring}, M.~T. {Eiles}, Z.~{Wadiasingh}, C.~A.
  {Taylor}, and C.~J. {Fitch}, {\it {Compton scattering in strong magnetic
  fields: Spin-dependent influences at the cyclotron resonance}},  {\em \prd}
  {\bf 90} (Aug., 2014) 043014, [\href{http://arxiv.org/abs/1408.2146}{{\tt
  arXiv:1408.2146}}].

\bibitem{1971AnPhy..67..599A}
S.~L. {Adler}, {\it {Photon splitting and photon dispersion in a strong
  magnetic field.}},  {\em Annals of Physics} {\bf 67} (1971) 599--647.

\bibitem{amego_2018_aa}
{AMEGO}, {\it {All Sky Medium Gamma-Ray Observatory}},  {\em
  \url{https://asd.gsfc.nasa.gov/amego/index.html}} (2019).

\bibitem{2018JHEAp..19....1D}
A.~{de Angelis}, V.~{Tatischeff}, I.~A. {Grenier}, J.~{McEnery},
  M.~{Mallamaci}, M.~{Tavani}, U.~{Oberlack}, L.~{Hanlon}, R.~{Walter},
  A.~{Argan}, and et~al., {\it {Science with e-ASTROGAM. A space mission for
  MeV-GeV gamma-ray astrophysics}},  {\em Journal of High Energy Astrophysics}
  {\bf 19} (Aug., 2018) 1--106, [\href{http://arxiv.org/abs/1711.01265}{{\tt
  arXiv:1711.01265}}].

\bibitem{2014APh....59...18H}
S.~D. {Hunter}, P.~F. {Bloser}, G.~O. {Depaola}, M.~P. {Dion}, G.~A. {DeNolfo},
  A.~{Hanu}, M.~{Iparraguirre}, J.~{Legere}, F.~{Longo}, M.~L. {McConnell},
  S.~F. {Nowicki}, J.~M. {Ryan}, S.~{Son}, and F.~W. {Stecker}, {\it {A pair
  production telescope for medium-energy gamma-ray polarimetry}},  {\em
  Astroparticle Physics} {\bf 59} (Jul, 2014) 18--28,
  [\href{http://arxiv.org/abs/1311.2059}{{\tt arXiv:1311.2059}}].

\bibitem{2018SPIE10699E..2MH}
S.~D. {Hunter}, {\it {The advanced energetic pair telescope for gamma-ray
  polarimetry}},  in {\em Space Telescopes and Instrumentation 2018:
  Ultraviolet to Gamma Ray}, vol.~10699 of {\em Society of Photo-Optical
  Instrumentation Engineers (SPIE) Conference Series}, p.~106992M, July, 2018.

\bibitem{gpst_ref}
{GPST}, {\it {The Gamma-ray Polarimetry Simulation Toolkit}},  {\em
  \url{https://github.com/ComPair/GPST}} (2019).

\end{thebibliography}\endgroup

\end{document}